\documentstyle[prl,aps,preprint]{revtex}
\tightenlines
\begin{document}
\title{Pseudogap in $\text{Bi}_2\text{Sr}_2\text{CaCu}_2\text{O}_{8+\delta}$ studied by measuring anisotropic susceptibilities and out-of-plane transport}
\author{Takao Watanabe,$^1$ Takenori Fujii,$^2$ and Azusa Matsuda$^{1,2}$}
\address{$^1$NTT Basic Research Laboratories, 3-1, Morinosato Wakamiya, Atsugi-Shi, Kanagawa 243-0198, Japan}
\address{$^2$Department of Applied Physics, Faculty of Science, Science University of Tokyo, 1-3 Kagurazaka, Shinjuku-ku, Tokyo 162-8601, Japan}
\date{\today}
\maketitle
\begin{abstract}
We find in the $\text{Bi}_2\text{Sr}_2\text{CaCu}_2\text{O}_{8+\delta}$ system that the characteristic temperatures $T^\ast_{\chi}$ (below which the uniform susceptibilities ${\chi}_{ab}(T)$ $(H{\perp}c)$ and ${\chi}_{c}(T)$ $(H{\parallel}c)$ decrease) and $T^\ast_{\rho_c}$ (below which the out-of-plane resistivity $\rho_{c}(T)$ shows typical upturn) coincide for all doping levels. We attribute the $T$ dependence of $\chi$'s and $\rho_{c}$ to the anomalous (pseudogapped) density-of-states (DOS) in high-$T_c$ cuprates. Furthermore, the anisotropy in the $T$ dependence of $\chi$'s is universal, i.e., $\chi_{c}\propto1.6\chi_{ab}$, showing that there is only a single $T$-dependent component in the $\chi$'s. This implies that the Curie-like behavior ($d\chi/dT<0$) observed in overdoped samples is also caused by a DOS effect.
\end{abstract}
\narrowtext
Although in recent years a lot of evidence for the pseudogap phenomenon in high-$T_c$ cuprates has been obtained, there is still no consensus about when (at what temperatures) it opens \cite{bat}. This is mainly because of our incomplete understanding of the $T$ dependence of the uniform susceptibility $\chi$. For example, in the $\text{La}_{2-x}\text{Sr}_x\text{CuO}_4$ system, it is known that $\chi$ first increases, takes a broad maximun, and then decreases from some characteristic temperature (here defined as $T^\ast_{\chi}$) with decreasing temperature and $T^\ast_{\chi}$ decreases with increasing doping (Sr content) \cite{john1}. D. C. Johnston has interpreted this as the sum of two-dimensional (2D) $S$=1/2 square-lattice Heisenberg antiferromagnetism and $T$-independent Pauli paramagnetism \cite{john1}. Recently, there has been proposed a model based on the mode-mode coupling theory of spin fluctuations in 2D metals with a technically nested Fermi surface, which explains the $T$ dependence as a crossover from the localized to itinerant spin fluctuations \cite{miya}. In those pictures, the gradual decrease in $\chi$ is not attributed to the formation of the pseudogap. On the other hand, J. W. Loram et al. have shown from susceptibility and high-resolution specific heat measurements that the pseudogap is developed below $T^\ast_{\chi}$ in the $\text{YBa}_2\text{Cu}_3\text{O}_{6+\delta}$ \cite{lor1} and $\text{La}_{2-x}\text{Sr}_x\text{CuO}_4$ systems \cite{lor2}. Similar results have been reported from NMR Knight shift measurements \cite{will1}. Up to now, we have not had a unified description of the $T$/doping dependence of $\chi$, that is the pseudogap-like behavior at low dopings and Curie-like behavior at high dopings.

Several characteristic temperatures are thought to indicate the opening of the pseudogap. In the case of $\text{YBa}_2\text{Cu}_3\text{O}_{6+\delta}$, $T^\ast_{\rho_c}$ coincides with $T^\ast_{\rho_a}$ \cite{tak} as well as $T^\ast_{\chi}$, where $T^\ast_{\rho_c}$ and $T^\ast_{\rho_a}$ are the temperatures below which out-of-plane resistivity $\rho_{c}(T)$ shows typical semiconductive upturn and in-plane resistivity $\rho_{a}(T)$ shows a deviation from high-temperature $T$-linear behavior. This supports the view that the characteristic change in $\rho_{a}(T)$, $\rho_{c}(T)$, and $\chi(T)$ comes solely from the opening of the pseudogap. However, in $\text{Bi}_2\text{Sr}_2\text{CaCu}_2\text{O}_{8+\delta}$, we have previously pointed out that $T^\ast_{\rho_c}$ and $T^\ast_{\rho_a}$ are different \cite{wat1}. Therefore, examining the above hypothesis with materials other than $\text{YBa}_2\text{Cu}_3\text{O}_{6+\delta}$ is important for understanding the meaning of the $T^\ast$'s. Recently, K. Ishida et al. have found from their NMR measurement \cite{ishi} that $T^\ast$ (the peak in 1/$T_1T$), $T^\ast_{K}$ (the steep decrease in the $^{63}$Cu Knight shift) and $T^\ast_{\rho_a}$ follow the same doping dependence. They have also noticed that $T_{mK}$ (the gradual decrease in the Knight shift from a constant and which thus corresponds to our $T^\ast_{\chi}$) follows a different doping dependence. They attribute $T^\ast_{K}$ to the pseudogap and $T_{mK}$ to a signature for the development of AF-spin correlations \cite{miya}. There has been up to now, however, no systematic study on the anisotropic susceptibilities ${\chi}_{ab}(T)$ $(H\perp{c})$, ${\chi}_{c}(T)$ $(H\parallel{c})$ and $\rho_{c}(T)$ for the $\text{Bi}_2\text{Sr}_2\text{CaCu}_2\text{O}_{8+\delta}$ system.

Single crystals were grown using the traveling solvent floating zone (TSFZ) method with a feed rod of $\text{Bi}_{2.1}\text{Sr}_{1.9}\text{CaCu}_2\text{O}_{8+\delta}$. To control  $\delta$ precisely, we first made the $P_{O2}-T$ equilibrium phase diagram of a $\text{Bi}_{2.1}\text{Sr}_{1.9}\text{CaCu}_2\text{O}_{8+\delta}$ single crystal for various $\delta$ using a thermogravimetric measurment. The samples were then annealed under the corresponding oxygen partial pressures at 600 $^\circ$C for 20$\sim$30 h, after which, in order to suppress high-temperature disorder that causes the Curie-Weiss behavior, they were slowly cooled to 300$\sim$400 $^\circ$C while keeping the equilibrium oxygen pressures. Finally, they were rapidly quenched \cite{wat1}. Heavily overdoped samples with $T_c\approx$60 K were made by high O$_2$ pressure (400 atm) annealing at 500 $^\circ$C for 50 h using a hot isostatic pressing (HIP) furnace. The $\delta$ for the HIP-treated sample was estimated as $\approx$0.3 from our equilibrium phase diagram \cite{wat1}. We attained high accuracy in the measurements by using the reciprocating sample option (RSO: absolute sensibility $\ge$5 x $10^{-8}$ emu) of a Quantum Design superconducting quantum interference device (SQUID) magnetometer as well as by using large crystals (sample weight $\approx$20 mg).

The magnetic susceptibilities ${\chi}_{c}(T)$ and ${\chi}_{ab}(T)$ under H=5 T of the $\text{Bi}_2\text{Sr}_2\text{CaCu}_2\text{O}_{8+\delta}$ single crystals with various oxygen contents ($\delta$) are shown in Figs. 1(a) and (b), respectively. The overall magnitude of ${\chi}_{c}(T)$ and ${\chi}_{ab}(T)$ increases with increasing $\delta$. We interpret this in terms of an increase in the density-of-states (DOS) near the Fermi level with carrier doping. The susceptibilities in the underdoped ($\delta$=0.22, $T_c$=82 K) and optimum-doped ($\delta$=0.25, $T_c$=89 K) samples monotonically decreases with decreasing temperature. For the slightly overdoped sample ($\delta$=0.26, $T_c$=86 K), they are almost constant at high temperatures and gradually decrease from a characteristic temperature $T^\ast_{\chi}$ [shown by the arrow in Fig. 1(a)] upon cooling. A further increase in $\delta$ causes a shift of $T^\ast_{\chi}$ to lower temperatures, whereas above $T^\ast_{\chi}$ the susceptibilities prominently decrease with increasing temperature. $T^\ast_{\chi}$ is estimated as 200, 175, and 135 K for $\delta$=0.26, 0.27, and 0.28, respectively. Here, $T^\ast_{\chi}$ was determined as the temperature at which $\chi_c$ deviates 2 \% from the high temperature $T$-linear behavior. The difference in the absolute value between $\chi_c$ and $\chi_{ab}$ may come from the anisotropic Van Vleck paramagnetism of the Cu$^{2+}$ ions, which is almost completely $T$-independent \cite{john2}.

The out-of-plane resistivity $\rho_{c}(T)$ for optimum and overdoped (0.26$\le$$\delta$$\le$0.3, 86 K$\ge$$T_c$$\ge$60 K) $\text{Bi}_2\text{Sr}_2\text{CaCu}_2\text{O}_{8+\delta}$ single crystals is shown in Fig. 2. The $\rho_{c}$ in a optimum-doped sample ($\delta$=0.25, $T_c$=89 K) shows semiconductive behavior in all temperature regions measured. With increasing $\delta$, the overall magnitude of $\rho_{c}$ decreases and the characteristic temperature $T^\ast_{\rho_c}$ (shown by the arrow in Fig. 2) for the onset of upturn appears. If $T^\ast_{\rho_c}$ is defined in the same manner as $T^\ast_{\chi}$, it can be estimated as 185, 165, and 130 K for $\delta$=0.26, 0.27, and 0.28, respectively. For the heavily overdoped sample ($\delta$$\approx$0.3, $T_c$=60 K), $\rho_{c}(T)$ shows no upturn behavior, suggesting the pseudogap completely vanishes at this doping level.

The above obtained $T^\ast_{\chi}$ and $T^\ast_{\rho_c}$ together with several other characteristic temperatures, including $T_c$, are plotted as a function of carrier concentration $p$ in Fig. 3. Here $p$ was estimated from $T_c$ using the empirical relation $T_c/T_{c,max}=1-82.6(p-0.16)^2$ \cite{tal}. $T^\ast_{\chi}$ and $T^\ast_{\rho_c}$ coincides for each $p$. Unlike $\text{YBa}_2\text{Cu}_3\text{O}_{6+\delta}$, which may change its dimensionality from two (2D) to three (3D) upon doping, the $\text{Bi}_2\text{Sr}_2\text{CaCu}_2\text{O}_{8+\delta}$ stays in 2D for all doping levels. We have previously shown that the $\rho_c$ in the $\text{Bi}_2\text{Sr}_2\text{CaCu}_2\text{O}_{8+\delta}$ system is governed by a simple tunneling process \cite{wat2}. Then $\rho_c$ should reflect the in-plane DOS. Since the magnitude of spin susceptibility also reflects the DOS, the coincidence of $T^\ast_{\chi}$ and $T^\ast_{\rho_c}$ simply indicates that the pseudogap opens at these temperatures. This is different from the explanation where the gradual decrease in $\chi$ is attributed to the development of AF-spin correlations. It should be noted, however, that the pseudogap behavior at the doping level of $\delta$=0.28 ($p$=0.193) may not be distinguishable from the superconductive DOS fluctuation effect \cite{var}. We have confirmed this pseudogap opening at $T^\ast_{\chi}$/$T^\ast_{\rho_c}$ by scanning tunneling spectroscopy (STS) (shown as $T^\ast_{tunnel}$ in Fig. 3) \cite{mat1,mat2}, although we recently recognized that there was a little shift in the doping level due to oxygen depletion from the surface in the experiment in \cite{mat1}. Our results here may conflict with the NMR measurement \cite{ishi} and the angle-resolved photoemission spectroscopy (ARPES) \cite{din}, which show that the pseudogap opens at around $T^\ast_{\rho_a}$,(shown as $T^\ast_K$ and $T^\ast_{ARPES}$, respectively, in Fig. 3). We could not observe any discontinuous change in temperature dependence or anisotropy of the susceptibility at around $T^\ast_{\rho_a}$. We point out, however, that if $T^\ast$ in \cite{ishi} is defined as a temperature at which 1/$T_1T$ departs from high-temperature Curie-Weiss behavior, it will come close to $T_{mK}$ and thus will come close to our pseudogap phase boundary. A slight dip structure in the DOS, which causes the decrease in the susceptibility, may not have been assigned as a gap opening by the ARPES in terms of the leading edge analysis. In order to understand the anomaly seen at $T^\ast_{\rho_a}$, one may need to consider the strongly k-dependent quasiparticle lifetime, or it may be simply unrelated to the pseudogap phenomena.

Next, in order to analyze the $T$ dependence of anisotropic susceptibilities quantitatively, we plotted ${\chi}_{c}(T)$ vs. ${\chi}_{ab}(T)$ with an implicit parameter $T$ for several doping levels (Fig. 4). Here, we assume that the $T$ dependence comes from spin susceptibilities of the doped CuO$_2$ plane. Each of the ${\chi}_{c}(T)$ vs. ${\chi}_{ab}(T)$ plots [Figs. 4(a), (b) and (c)] shows the same linear relation, i. e., $\chi_{c}\propto1.6\chi_{ab}$, except near $T_c$ ($\le$140 K). This implies that the spin susceptibility has a single component with the ratio of anisotropic $g$-factors, $(g_c/g_a)^2$=1.6, and thus the Curie-like explanation for the ${\chi}$ of overdoped samples may be excluded. This is because it is unlikely that the localized paramagnetic centers, which could have appeared additively in the overdoped state, have accidentally the same anisotropy as the main spin component. This single-component hypothesis is consistent with NMR experiments \cite{all} and ensures that we are actually measuring the intrinsic spin susceptibility. In Fig. 3, we included the temperatures at which the deviation from the linear scaling becomes evident, $T_{scf}$. Since the $T_{scf}$ scales with $T_c$, it is considered to be the onset temperature of Aslamasov-Larkin (AL)-type superconductive fluctuation, which is extremely anisotropic in high-$T_c$ cuprates \cite{john2}.

As already mentioned, we consider that the $T$ dependence of $\chi$ for underdoped samples is caused by the pseudogap opening, which has been evidenced by STS \cite{mat1,mat2,ren,miyak}. Then, it is natural to consider that the Curie-like behavior of $\chi$ for overdoped samples is also caused by a DOS effect, since the system is expected to be more Fermi-liquid like upon doping. One such possibility is an van-Hove singularity (vHs), whose existence has been confirmed by ARPES at somewhat below the Fermi energy ($\varepsilon_F$) for a slightly overdoped sample \cite{she}. The vHs is considered to grow and come near the $\varepsilon_F$ with increasing doping. Here, we assume simply the Pauli paramagnetism for the single-component spin susceptibility, $\chi_{pauli}=\mu_B^2$${\int}_{-\infty}^{\infty}N(\varepsilon)(-{\partial}f(\varepsilon)/\partial\varepsilon)d\varepsilon$, where $N(\varepsilon)$ is the DOS and $f(\varepsilon)$ is the Fermi function. Since $\chi_{pauli}$ reflects the thermally averaged DOS near $\varepsilon_F$, when one takes into account the anomalous energy-dependent DOS (pseudogap and vHs) and its doping evolution, it will explain the $\chi$ observed. This kind of interpretation was first proposed by J. W. Loram et al. \cite{lor2} and later asserted by G. V. M. Williams et al. \cite{will2}. Our data here support their interpretation from a different point of view.

The assumption of the Pauli paramagnetism for the spin susceptibility enables us to estimate the pseudogap value. Here in the underdoped state, the energy independent background DOS, $N_0$, was assumed (the vHs may sit far below $\varepsilon_F$ and be smeared out). For the pseudogap, the BCS-type DOS (Dynes formula) with the d-wave symmetry gap was assumed, $N(\varepsilon)/N_0=\int_0^{2\pi}(d\theta/2\pi)Re[\mid\varepsilon-i{\Gamma}\mid/\sqrt{(\varepsilon-i\Gamma)^2-(\Delta\cos(2\theta))^2}]$, where $\Gamma$ is the quasiparticle scattering rate, which was assumed here as 2$k_B$T in a consistent way with the transport data, and $\Delta$ is a maximum d-wave gap, the temperature dependence of which was neglected. Then we can fit the experimental spin susceptibility $\chi^{iso}_{spin}(T)$ \cite{the} by the calculated ones with a free parameter $\Delta$ as shown in Fig. 5. The pseudogap value $\Delta$ was estimated as 65$\pm$3 and 48$\pm$3 meV for the underdoped ($\delta$=0.22) and optimum doped ($\delta$=0.25) samples, respectively. These pseudogap values coincide with our direct STS measurements \cite{mat1,mat2} and, possibly, with the "high energy gap" recently suggested by photoemission spectroscopy \cite{ron}, but are larger than the superconducting gap (20 meV$\le$$\Delta$$\le$45 meV) obtained by ARPES/STS \cite{mat1,mat2,din,ren,miyak}.

In conclusion, the experimentally obtained pseudogap phase boundary ($T^\ast_{\chi}$ and $T^\ast_{\rho_c}$) is not a smooth extrapolation of the $T_c$ boundary of the heavily overdoped state, rather, it seems to cross the $T_c$ boundary at a slightly overdoped level (Fig. 3). This result together with the estimated larger pseudogap compared to superconducting gap suggests the two gaps are different in origin, although there are many arguments that the pseudogap is a precursor of the d-wave superconducting gap \cite{ren}.

We would like to thank Professors M. Kohmoto and M. Suzuki for helpfull discussions and Dr. H. Shibata for extending the use of the HIP facility.

\vfill
\eject
\noindent
Figure captions
\par
\noindent
\\
Fig. 1.
Magnetic susceptibilities (a) ${\chi}_{c}(T)$ $(H\parallel{c})$ and (b) ${\chi}_{ab}(T)$ $(H\perp{c})$ of the $\text{Bi}_2\text{Sr}_2\text{CaCu}_2\text{O}_{8+\delta}$ single crystals with various oxygen contents ($\delta$) under H=5 T. The solid straight lines in (a), which are linear extraporations of $\chi_c$ at higher temperatures, are eye guides for the overdoped ($\delta$=0.26, 0.27, and 0.28) samples. Arrows indicate the temperatures $T^\ast_{\chi}$ at which $\chi_c$ starts to decrease from its linear high temperature behaviors.
\par
\noindent
Fig. 2.
Out-of-plane resistivity $\rho_{c}(T)$ of a $\text{Bi}_2\text{Sr}_2\text{CaCu}_2\text{O}_{8+\delta}$ single crystal for oxygen content 0.26$\le$$\delta$$\le$0.3. The solid straight lines, which are linear extraporations of $\rho_{c}$ at higher temperatures, are eye guides for the overdoped ($\delta$=0.26, 0.27, and 0.28) samples. Arrows indicate the temperatures $T^\ast_{\rho_c}$ below which $\rho_{c}$ shows typical upturn.
\par
\noindent
Fig. 3.
Characteristic temperatures obtained by various measurements versus carrier concentration $p$. The $p$ values (including for the previous results) were systematically estimated from the empirical $T_c$ vs. $p$ relation in the text.
\par
\noindent
Fig. 4.
${\chi}_{c}(T)$ versus ${\chi}_{ab}(T)$ plots for (a) underdoped ($\delta$=0.22), (b) slightly overdoped ($\delta$=0.26) and (c) heavily overdoped ($\delta$$\approx$0.3) $\text{Bi}_2\text{Sr}_2\text{CaCu}_2\text{O}_{8+\delta}$ single crystals. Some typical temperatures are shown by arrows.
\par
\noindent
Fig. 5.
Isotropic component of the spin susceptibilities, $\chi^{iso}_{spin}(T)$, of a $\text{Bi}_2\text{Sr}_2\text{CaCu}_2\text{O}_{8+\delta}$ single crystal for oxygen content $\delta$=0.22 (underdoped) and 0.25 (optimum doped). The solid lines are numerical fits assuming the BCS-type DOS for the pseudogap.
\\


\begin{references}
\bibitem{the}The isotropic component of the spin susceptibilities, $\chi^{iso}_{spin}$, was obtained by averaging the anisotropic spin susceptibilities, i. e., $\chi^{iso}_{spin}$=(2/3)$\chi^{ab}_{spin}$+(1/3)$\chi^{c}_{spin}$, where $\chi^{ab}_{spin}$ and $\chi^{c}_{spin}$ were obtained using the values in the literature for $\chi_{ab}^{VV}$ and $\chi_{c}^{VV}$ \cite{john2}.
\end{references}
\end{document}